\documentclass[aps,pra,epsfigure,twocolumn,showpacs]{revtex4}
\usepackage{dcolumn}
\usepackage{bm}
\usepackage{graphicx}
\usepackage{amsmath}
\usepackage{latexsym}
\usepackage{amsfonts}
\usepackage{amssymb}
\usepackage{array}
\usepackage{epsfig}
\usepackage{epstopdf}

\newcommand{\ket}[1]{\left\vert#1\right\rangle}

\newcommand{\bra}[1]{\left\langle#1\right\vert}

\newcommand{\beq}{\begin{equation}}
\newcommand{\eeq}{\end{equation}}
\newcommand{\bea}{\begin{eqnarray}}
\newcommand{\eea}{\end{eqnarray}}
\newcommand{\noon}{\textrm{NOON}}

\begin{document}
\title{Transferring entanglement to the steady-state of flying qubits}
\author{Yanqiang Guo$^1$, Jie Li$^2$, Tiancai Zhang$^1$, and Mauro Paternostro$^{2,3}$}
\affiliation{$^1$State Key Laboratory of Quantum Optics and Quantum Optics Devices, Institute of Opto-Electronics, Shanxi University, Taiyuan 030006, China\\
$^2$Centre for Theoretical Atomic, Molecular and Optical Physics, School of Mathematics and Physics, Queen's University, Belfast BT7 1NN, United Kingdom\\
$^3$Institut f\"ur Theoretische Physik, Albert-Einstein-Allee 11, Universit\" at Ulm, D-89069 Ulm}

\begin{abstract}
The transfer of entanglement from optical fields to qubits provides a viable approach to entangling remote qubits in a quantum network. In cavity quantum electrodynamics, the scheme relies on the interaction between a photonic resource and two stationary intra-cavity atomic qubits. However, it might be hard in practice to trap two atoms simultaneously and synchronize their coupling to the cavities. To address this point, we propose and study entanglement transfer from cavities driven by an entangled external field to controlled flying qubits. We consider two exemplary non-Gaussian driving fields: NOON and entangled coherent states. We show that in the limit of long coherence time of the cavity fields, when the dynamics is approximately unitary, entanglement is transferred from the driving field to two atomic qubits  that cross the cavities. On the other hand, a dissipation-dominated dynamics leads to very weakly quantum-correlated atomic systems, as witnessed by vanishing quantum discord. 
\end{abstract}
\date{\today}
\maketitle

\section{Introduction}

Entanglement is an invaluable resource for many important tasks in quantum information processing~\cite{Nielsen,Bennett,Duan,Wineland,Kimble}, including metrological purposes~\cite{Roos,Riedel}. Schemes for the generation and distribution of entanglement have been designed and implemented, in the past 20 years, in a number of physical systems. The quality of these strategies has considerably increased over recent years, making the creation of multipartite entangled states of a few elements a reality~\cite{Pan,Blatt,Martini} and paving the way to the near-future realization of networks of distributed quantum nodes for quantum communication and computing~\cite{Kimble}. The main challenge, in this context, is the achievement of reliable interfaces between information carriers having different natures.

This problem has long been investigated, both theoretically and experimentally, and various solutions for the achievement of controllable interactions between static local processors and flying information carriers have been designed~\cite{Kimble}. Among them, a promising one for its technologically realistic nature and its flexibility is embodied by the transfer of quantum correlations from light to matterlike systems~\cite{Kraus,Paternostro,Paternostro2004,Zou,LeeBose,ACIP,Paternostro2010}.
Broadly speaking, this paradigm for entanglement distribution via light-matter interfaces requires the availability of entangled-light resources and the ability to perform local light-matter interactions that {\it pass} the quantum correlations (or part of them) to initially separable local matterlike systems. Originally devised for cavity and circuit QED settings~\cite{Kraus,Paternostro,Paternostro2004}, this approach has recently been extended to mechanical systems interfaced to light~\cite{LauraMauroJie} and quantum many-body systems~\cite{zippilli}.

In most of the above-mentioned schemes based on cavity-QED technology, it has been assumed that multi-level atoms are located at a fixed point within a cavity, where they  are coupled to an antinode of the cavity-field standing wave. In the actual situation, the atoms are either flying through a cavity~\cite{Zhang} or trapped within an intra-cavity optical dipole trap, yet still moving within it~\cite{Hood}. Current records for atomic optical traps reach trapping times as long as 30 s~\cite{Hijlkema}. However, the management of a network of single atoms trapped in distant cavities embodies a considerable challenge. Moreover, although efficient ways to switch the cavity-atom interaction exist, such a configuration makes it hard to arrest the evolution of the inter-atom entanglement so as to achieve a desired value.

In this paper, we propose a simple strategy to bypass both such difficulties. We study entanglement transfer to two flying atomic qubits crossing two remote cavities that are driven by an entangled resource. 
The scheme relies on the high level of synchronization that can be arranged for the passage of two different atoms across remote  cavities. Moreover, as the light-atom interaction is turned off after the atoms exit the cavities, it is possible to arrange for steady-state atomic entanglement.

As entangled resources, we consider both NOON states and entangled coherent states~\cite{sanders} (ECSs), which are important non-Gaussian resources of experimental relevance (five-photon NOON states have been produced recently~\cite{Afek}, while ECSs with an amplitude of about one photon can be produced, using a beam splitter, from the coherent superpositions of coherent states described in Ref.~\cite{Jeong}). Both classes of states have broad prospective applications in either quantum metrology \cite{Lee} or fundamental research \cite{Kok}. We study the efficiency of the entanglement-transfer scheme in various dynamical regimes, highlighting the effectiveness of steady-state entanglement distribution in the quasi-coherent case corresponding to the good-cavity limit, while stating a {\it no-go} result for the regime dominated by cavity dissipation.

The remainder of the paper is structured as follows. In Sec.~\ref{Model} we describe the system we address and describe the preparation of the entangled cavity fields. Section~\ref{results} is devoted to the analysis of the relative performance of the scheme under the two classes of entangled resources. In Sec.~\ref{impossible} we address the dissipation-dominated dynamics, showing the impossibility of entanglement transfer. Finally, we draw our conclusions in Sec.~\ref{Con}.

\begin{figure}[b]
\includegraphics[width=0.95\linewidth]{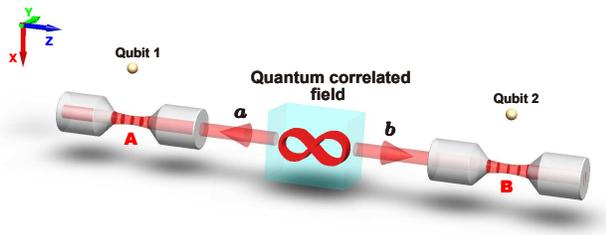}
\caption{(Color online) Scheme of the protocol: Two remote and identical cavities are driven by a two-mode quantum correlated field, which is coupled to each cavity via a leaky mirror. Two qubits are then sent to pass simultaneously through the cavities along the vertical ($x$) direction from the tops of the geometrical centers.}
\label{model}
\end{figure}

\section{The model}
\label{Model}

Let us introduce here the  scheme that we address in this work, which is shown in Fig.~\ref{model}. Two freely propagating field modes, labeled in the following $a$ and $b$, drive two remote single-mode cavities, which we call $A$ and $B$, respectively. The state of the driving fields will be specified, when needed, later on. The cavities are prepared in their vacuum state. The coupling between each external driving field and the corresponding cavity is modeled as a beam splitter (BS) with transmittivity (reflectivity) $T$ ($R=1-T$)~\cite{Kim1995} that is quantitatively determined by the quality factor of each cavity~\cite{Paternostro2004}. 
The state of the cavity modes that result from the driving process is given by
 \begin{equation}
\rho_{AB}(0)=\text{Tr}_{ab}[\hat B_{Aa}\hat B_{Bb} \rho_{abAB}(0)\hat B^\dag_{Aa}\hat B^\dag_{Bb}],
\label{eq4}
\end{equation}
where $\hat B_{Aa}$ and $\hat B_{Bb}$ are the BS operators. As a result, in general, the cavity fields become entangled. Two two-level atoms (qubits) with energy eigenstates $|0\rangle_i$ and $|1\rangle_i$ ($i{=}1,2$) and Bohr frequencies equal to the frequency of the cavity fields pass through their respective cavities. Here we consider two cases: {\bf (a)} the simultaneous free-falling of the atoms across the  cavity (i.e., the qubits are accelerated by gravity)~\cite{Zhang}; {\bf (b)} the constant-speed passage of the atoms, which can be realized experimentally by embedding an optical lattice loaded with a single atom into the cavity~\cite{Hijlkema,Kuhr} or using optical traps for atomic confinement and transport. For short time intervals, the coupling strength between each atom and the corresponding cavity  field can be considered as a constant, and the local qubit-field interaction can be treated through the standard (resonant) Jaynes-Cummings (JC) model that reads, in the interaction picture with respect to the free energy of qubits and cavities, as 
$\hat H_{jk_j}(t){=}\hbar\Omega(x(t),y(t))(\hat k_j|1\rangle_j\langle0|+\hat k^\dag_j|0\rangle_j\langle1|)$ (with $j{=}1,2$ and $k_1{=}A,k_2{=}B$). Here,
$\Omega(x(t),y(t))$ is the qubit-field coupling frequency, which depends on the position of an atom within the respective cavity and the form of the field mode. As a simplifying assumption that does not affect the generality of our analysis, we consider the atoms falling perfectly vertically. This means assuming $y{=}0$ in our model (we take the origin of each reference frame at the geometrical center of the respective cavity). For a transverse Hermite-Gauss mode, $\Omega(x(t),0)$ takes the form~\cite{Rempe}
\begin{equation}
\Omega(x(t),0)=\Omega_0\frac{|\Psi_{u,v}(x(t),0)|}{|\Psi_{0,0}(0,0)|},
\label{eq7}
\end{equation}
where $\Omega_0$ is a constant and $\Psi_{u,v}(x(t),0)$ is the mode function of the cavity,
\begin{equation}
\Psi_{u,v}(x(t),0)=C_{u,v} e^{{-}\frac{x^2(t)}{\omega^2_0}}H_u\left(\frac{\sqrt{2}x(t)}{\omega_0}\right).
\label{eq8}
\end{equation}
Here, we have introduced the coefficient $C^{-1}_{u,v}=(2^u2^vu!v!)^{1/2}(\omega_0^2\pi/2)^{1/2}$, while $H_{p}(x)$ is the Hermite polynomial of order $p$ and argument $x$. The waist $\omega_0$ of each field mode is determined by the radius of curvature of the mirrors and the cavity length~\cite{Zhang}. 

Let us first assess the unitary dynamics resulting from neglecting any source of noise and losses. The associated unitary operator is $\hat U_{12AB}(t)=\hat U_{1A}(t){\otimes}\hat U_{2B}(t)$, where $\hat U_{jk_j}(t)=\exp[-i\int^t_0\hat H_{jk_j}(t')dt'/\hbar]$. The dynamics of the two flying qubits is, however, nonunitary and described by the reduced density matrix 
\begin{equation}
\rho_{12}(t)=\text{Tr}_{AB}[\hat U_{12AB}(t)\rho_{12}(0)\otimes\rho_{AB}(0)\hat U^{\dag}_{12AB}(t)].
\label{eq9}
\end{equation}
In order to fix the ideas, we assume that the qubits are initialized in $|00\rangle_{12}\langle00|$, which is usually the preparation for which the entanglement-transfer process is optimized~\cite{MP}.

As for the entangled resource, in this paper we will concentrate on two families of non-Gaussian states of experimental significance, namely, the NOON and ECS families. The first family is described as
\begin{equation}
|\noon\rangle_{ab}=\frac{1}{\sqrt{2}}(|\textrm{N0}\rangle_{ab}+|\textrm{0N}\rangle_{ab})~~(\textrm{N}\in\mathbb{N}),
\label{eq1}
\end{equation}
while the ECS representative that will be used in our study is
\begin{equation}
\left\vert\mathrm{ECS}\right\rangle _{ab}=\mathcal{N}_{\alpha}(\left\vert
\alpha0\right\rangle _{ab}{+}\left\vert 0\alpha\right\rangle _{ab}),
\label{eq13}
\end{equation}
where $\mathcal{N}_{\alpha}=[2(1{+}e^{-\vert \alpha\vert
^{2}})]^{-1/2}$ is the state normalization. In what follows, we study quantitatively the entanglement-transfer scheme performed using each of such states.

\section{Performance of the entanglement-transfer process}
\label{results}

Let us begin with the NOON family, so that the initial state of the pump-cavity system is $|\noon\rangle_{ab}{\otimes}|00\rangle_{AB}$. The field that is thus prepared within the cavities is~ \cite{Li}
 \begin{equation}
 \begin{split}
 \rho_{AB}=&\frac{1}{2}\sum^N_{m=0}\begin{pmatrix}N\\
m\end{pmatrix}R^{N-m}T^m(|m0\rangle_{AB}\langle m0|+|0m\rangle_{AB}\langle 0m|)\\
 &+\frac{T^N}{2}(|N0\rangle_{AB}\langle 0N|+|0N\rangle_{AB}\langle N0|).
\end{split}
 \label{eq5}
 \end{equation}
An interesting point that will be reprised and explained later on in this paper is that for $N>1$ the off-diagonal elements of the two-qubit density matrix are all exactly null, so that no entanglement is transferred. On the other hand, for $N=1$, the evolved atomic state takes the form (in the ordered basis $\{|11\rangle, |10\rangle,|01\rangle,|00\rangle\}_{12}$)
\begin{equation}
\rho_{12}(t)=
\begin{pmatrix}
0 & 0 & 0 & 0 \\
0 & TS/2 & TS/2  & 0 \\
0 & TS/2  & TS/2  & 0 \\
0 & 0 & 0 & 1-TS \\
\end{pmatrix},
\label{eq10}
\end{equation}
where $S=\sin^2[\mu(t)]$ and $\mu(t)=\int^t_0\Omega(x(t'),0)dt'$.  It should be noted that the impossibility of entanglement transfer for $N>1$ is not related to the fact that a single-photon entangled state violates a Clauser-Horne Bell inequality more than any other NOON state with $N>1$~\cite{Wildfeuer2007} but to the intrinsic structure of correlations in the resources at hand and the entanglement-transfer process ruled by the JC model.

We now consider another state of the driving field, namely, an entangled coherent state~\cite{sanders}, which is known to offer advantages over the NOON class. For instance, in some cases ECSs show a remarkably improved sensitivity for phase estimation as compared to that of the NOON states~\cite{Joo}. Such superiority, as we will see, does not extend to the entanglement-transfer paradigm performed using the JC model for the local atom-field interaction, as ECSs exhibit lower entangling power than NOON states.

An observation that allows for the understanding of the depleted entangling capacity of ECSs comes from realizing that the latter can be seen as a superposition of NOON states as~\cite{Afek,Luis2001}
\begin{equation}
\left\vert\mathrm{ECS}\right\rangle _{ab}=\mathcal{N}_{\alpha}e^{-\left\vert
\alpha\right\vert ^{2}/2}\sum_{n=0}^{\infty}\frac{\alpha^{n}}{\sqrt{n!}%
}(\left\vert n0\right\rangle_{ab}{+}\left\vert
0n\right\rangle _{ab}).
\label{eq15}
\end{equation}
It is straightforward to realize that the incommensurate nature of the Rabi frequencies $\Omega(x(t),0)\sqrt{n}$ at which entanglement is transferred to each subspace spanned by $\{\ket{n-1,1},\ket{n,0}\}_{k_j,j}$ generates quantum interference effects. The result is that the amount of entanglement effectively passed from the ECS resource to the qubit receivers is less than what is achieved using an entangled single-photon state.
In order to see this, we have quantified the entanglement between the flying qubits (initially prepared in their ground state), after their interaction with the driven cavities. The reduced two-bit density matrix is
\begin{equation}
\rho_{12}(t)=
\begin{pmatrix}
0 & 0 & 0 & 0 \\
0 & A & B & C \\
0 & B & A & C \\
0 & C^* & C^* & 1-2A
\end{pmatrix},
\label{eq18}
\end{equation}
with
\begin{equation}
\begin{aligned}
& A=\tilde{\mathcal{N}}_{\alpha}^{2}\sum_{n=1}^{\infty}\frac{\alpha^{2n}}{n!}\sum_{m=0}^{n}({\cal C}^n_m)^2R^{m}T^{n-m}\sin^{2}[\mu(t)\sqrt{n{-}m}],\\
& B=\tilde{\mathcal{N}}_{\alpha}^{2}T\alpha
^{2}\sin^{2}[\mu(t)],\\
& C=-i\tilde{\mathcal{N}}_{\alpha}^{2}\{2\sqrt
{T}\alpha\sin[\mu(t)]+\sum^\infty_{s=1}\sum^s_{m=0}\frac{{\alpha^{2s+1}}R^mT^{s-m+1/2}}{m!\sqrt{(s+1-m)!(s-m)!}}\\
&\times\sin[\mu(t)\sqrt{s-m+1}]\cos[\mu(t)\sqrt{s-m}]\}
\end{aligned}
\label{eq19}
\end{equation}
and $\tilde{\mathcal{N}}_{\alpha}=\mathcal{N}_{\alpha}e^{-|\alpha|^2/2}$,  ${\cal C}_{m}^{n}=\sqrt{n!/[m!(n-m)!]}$.

In order to quantify the qubit entanglement, in the remainder of this paper we use the negativity of $\rho_{12}(t)$~\cite{Vidal2002},
\begin{equation}
{\cal N}[\rho_{12}(t)]=\max[0,-2\lambda_-(t)],
\label{eq11}
\end{equation}
where $\lambda_-(t)$ is the smallest eigenvalue (with its sign) of the partially transposed two-qubit density matrix.

\subsection{Discussion of the results}

\begin{figure*}[t]
(a)\hskip5.6cm(b)\hskip5.6cm(c)
\includegraphics[width=0.33\linewidth]{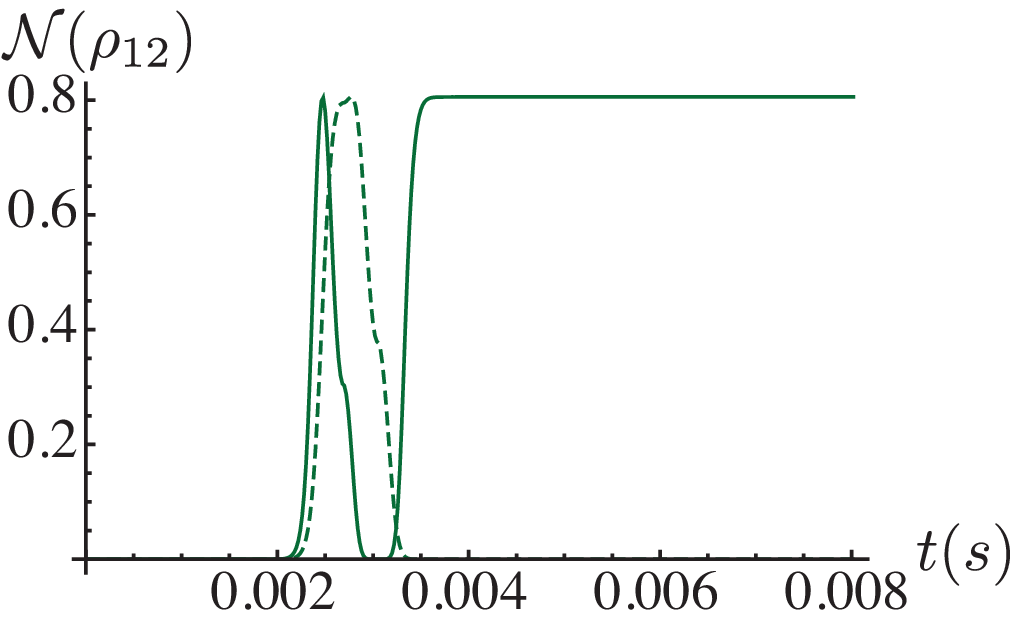}~~~\includegraphics[width=0.33\linewidth]{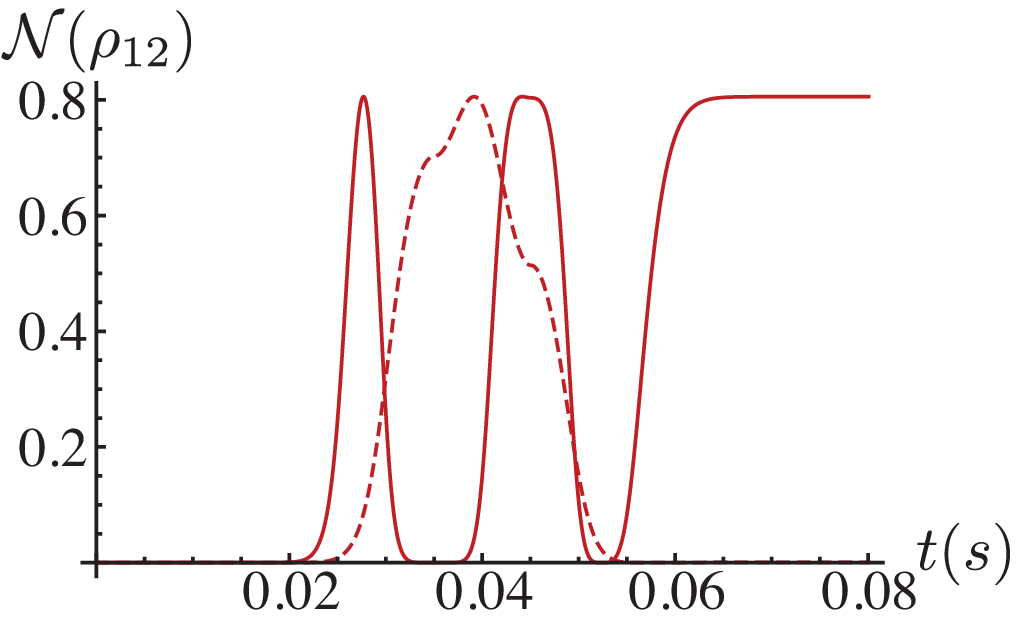}~~~\includegraphics[width=0.33\linewidth]{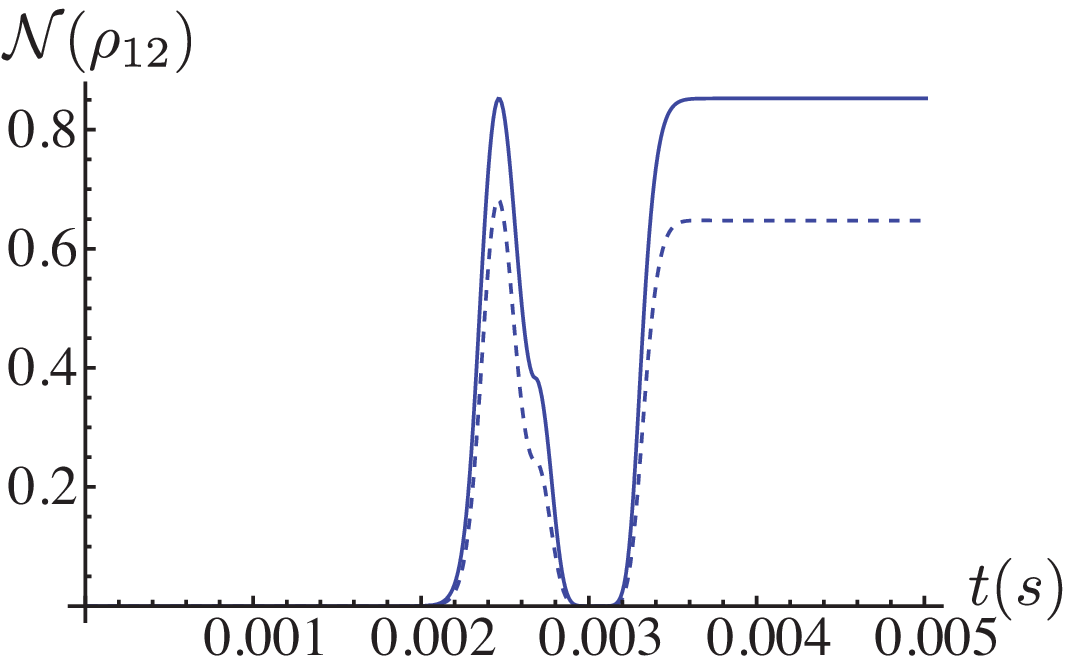}
\caption{(Color online) (a) Entanglement transfer for free-falling atoms and a single-photon entangled driving field (i.e., NOON states with $N=1$), $x(t)=x_0{+}gt^2/2$, where $x_0=-4\omega_0$, $g=9.82$ m/s$^{2}$, and for $\Omega_0=4$ kHz (dashed line); $\Omega_0=5.9$ kHz (solid line). (b) Entanglement of atomic qubits crossing the cavities under conditions of uniform motion [$x(t)=x_0{+}Vt$ with $x_0=-4\omega_0$ and $V=0.001$ m/s] and with the same driving field used in (a). We have taken $\Omega_0=155$ Hz (dashed line) and $\Omega_0=365$ Hz (solid line). (c) Comparison between the unitary (solid line) and dissipation-affected (dashed line) entanglement-transfer performance for free-falling motion of the atomic qubits with $\Omega_0=5.9$ kHz. In the dissipation-affected simulation, we have taken $\Omega_0/\Gamma\simeq10^{2}$. In all our calculations we have taken the cavity mode $\Psi_{2,0}(x,0)$ (i.e., the $\text{TEM}_{20}$ mode), $\omega_0=10$ $\mu$m, and $T=0.9$, and the qubits are prepared in $|00\rangle_{12}$.}
\label{fig2}
\end{figure*}
\begin{figure*}[t]
(a)\hskip5.6cm(b)\hskip5.6cm(c)
\includegraphics[width=0.33\linewidth]{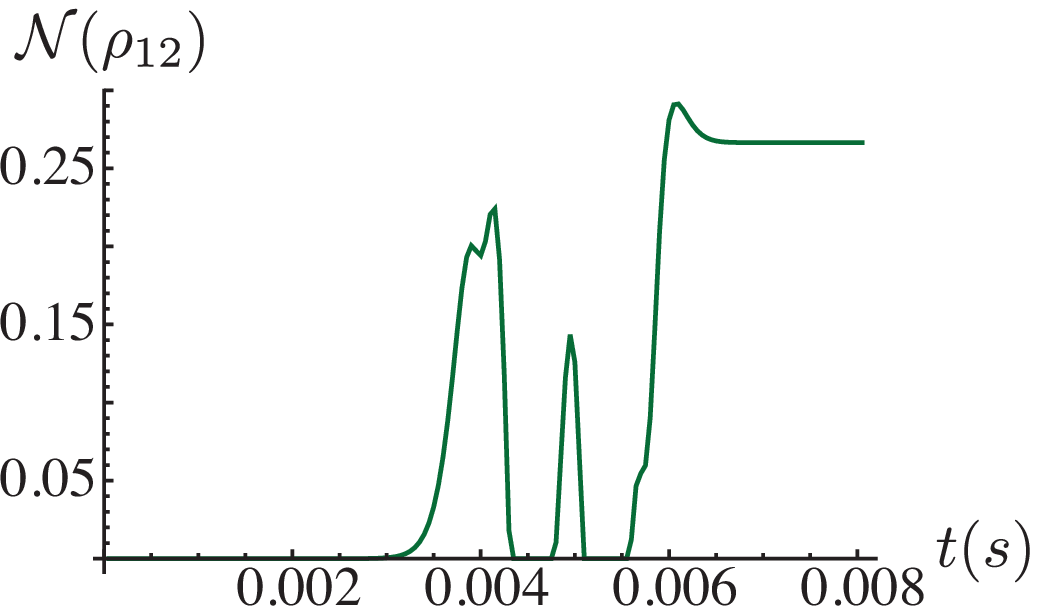}~~
\includegraphics[width=0.33\linewidth]{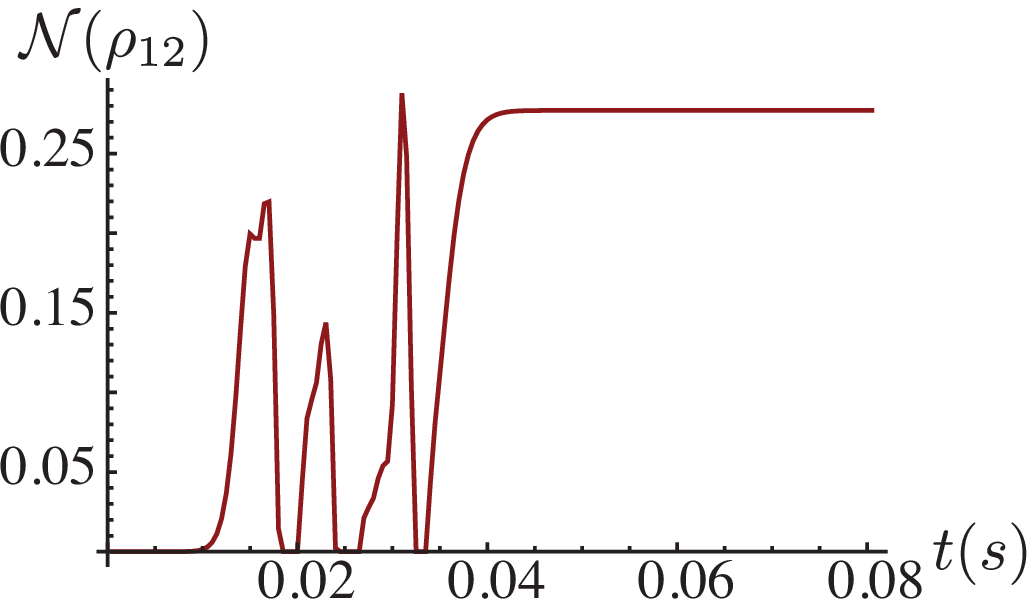}~~
\includegraphics[width=0.33\linewidth]{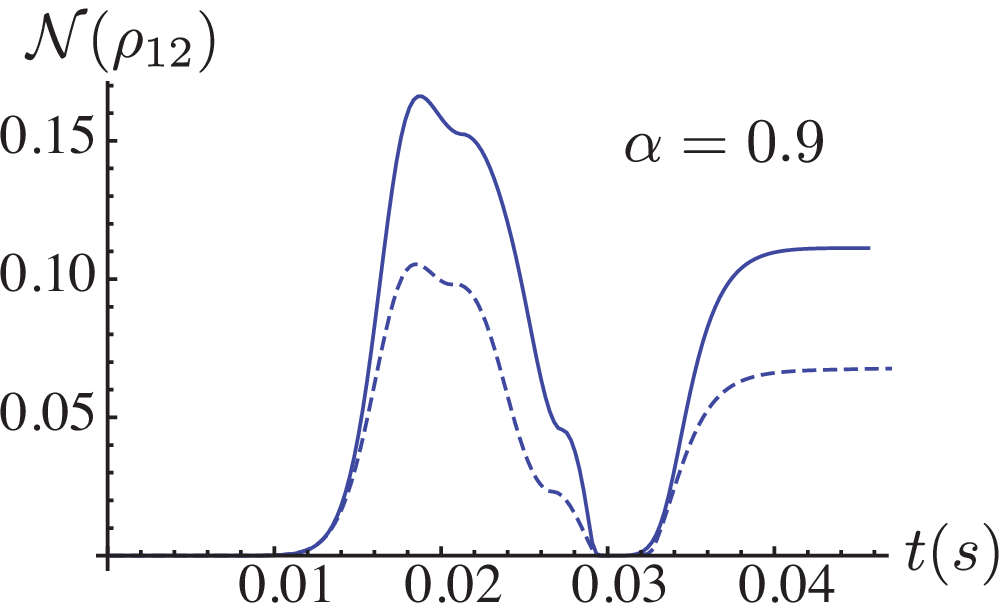}~~
\caption{(Color online) (a) Entanglement transfer for free-falling atoms and ECS external driving with $x_0=-4\omega_0$, $\alpha{=}1.1$, and $\Omega_0=6.1$ kHz. (b) Entanglement for atomic qubits crossing the cavities under uniform motion ($x_0=-4\omega_0$, $V=0.005$ m/s) and $\Omega_0=850$ Hz, under the same driving as in (a). The atomic steady-state entanglement at the exit of the cavities for both motions is comparable. (c) Comparison between the unitary (solid line) and dissipation-affected (dashed line) entanglement-transfer behavior for uniform motion of the atomic qubits ($V=0.005$ m/s), $\Omega_0=340$ Hz, and $\alpha=0.9$. We have taken $\Omega_0/\Gamma\simeq10^{2}$. For all plots, we take the cavity mode $\Psi_{2,0}(x,0)$, $\omega_0=30$ $\mu$m, $T=0.9$, and initial atomic ground state.}
\label{fig4}
\end{figure*}

The behavior of the qubit entanglement under a NOON and an ECS driving and a perfectly unitary dynamics is shown in Figs.~\ref{fig2}(a) and~\ref{fig2}(b) and~\ref{fig4}(a) and~\ref{fig4}(b), respectively. Clearly, a single-photon entangled state is able to prepare a sizably entangled qubit state depending on the value of the frequency $\Omega_0$. If the latter is such that the atoms perform an odd half-integer (integer) number of Rabi floppings by the time they cross the field modes, the flying qubits leave the cavities with a significant degree of steady-state entanglement (in a separable state), regardless of the motion (whether at constant velocity or accelerated) that they underwent within the resonators [cf. Figs.~\ref{fig2}(a) and~\ref{fig2}(b)]. The working conditions of the process can be adjusted so that the values of the steady-state entanglement achieved in the two instances of atomic motion are comparable (cf. Fig.~\ref{fig2}). Needless to say, for free-falling qubits, a Rabi frequency much larger than the one that should be chosen under conditions of constant velocity is required, as the interaction time between atoms and cavity fields is shortened.
As anticipated, the use of an ECS driving does not result in an equally effective transfer efficiency. Figures~\ref{fig4}(a) and~\ref{fig4}(b) show the negativity for both free-falling and constant-velocity qubits, optimized numerically over the parameters of the interaction and the amplitude of the ECS state (at $\alpha{\simeq}1.1$ the entanglement between the flying qubits is maximum, uniformly with respect to the choice of the other parameters of the model).

As seen from the analysis above, interaction times of the order of $10^{-3}$ (for accelerated qubits) or $10^{-2}$ (for atoms moving at constant velocity) are needed in order to cross the cavities. These values of the interaction time might put the dynamics of the system well within the typical time scale of leakage of the cavity fields due to finite cavity quality factors. Therefore, in order to make our analysis consistent with a realistic experimental situation, we should abandon the perfectly unitary description adopted so far in favor of an open-system dynamics that includes the effect of field dissipation. This is done, in what follows, by implementing a quantum Monte Carlo unraveling of the system's dynamics~\cite{MC} implemented by letting the system evolve through the non-Hermitian operator
\begin{equation}
\hat{\cal H}(t)=\sum^2_{j=1}\hat{H}_{jk_j}(t)-i\Gamma\hat{k}^\dag_j\hat{k}_j~(k_1=A, k_2=B)
\end{equation}
with $\Gamma$ the damping rate of each field (assumed to be the same for both the cavities, an assumption that can be easily relaxed if needed) and considering the effects of the quantum jump operators $\sqrt{\Gamma}\hat{k}_j$. The results of our numerical simulations of a large number of possible dynamical histories of the system are presented in Figs.~\ref{fig2}(c) and~\ref{fig4}(c) for $\max_t\Omega(x(t),0)/\Gamma\sim10^{2}$, which is a regime that can be achieved experimentally~\cite{Zhang,guopaper} [in Fig.~\ref{fig4}(c) we address the suboptimal case of $\alpha=0.9$ simply for convenience of calculations in light of the size of the truncated Hilbert space within which we have performed our quantum-jump calculations]. The process shows a noticeable robustness to the effects of field damping, leaving the qubits at the output of the cavities well entangled, irrespective of the details of the atomic motion and the form of the driving field.

\section{Impossibility of entanglement transfer in dissipation-dominated regimes}\label{impossible}

Here we address the case of a dissipation-dominated dynamics of the cavity fields to demonstrate that no entanglement can be transferred in this case, regardless of the flying or trapped nature of the atomic qubits. Moreover, besides determining the regime for effective entanglement transfer, this approach will help us understanding some of the results that have been gathered in the previous sections. Let us start by introducing damping of the cavity fields into the dynamical evolution of the system, which will now be ruled by the master equation
\begin{equation}
\begin{aligned}
\partial_t\rho_{12AB}&=-i[\sum_{j=1,2}\hat H_{jk_j},\rho_{12AB}]+\sum_{j=A,B}\hat{\cal L}_j(\rho_{12AB})\\
&=(\hat{\cal L}_0+\hat{\cal L})(\rho_{12AB})
\end{aligned}
\end{equation}
with $\rho_{12AB}$ the density matrix of the whole cavity-qubit system and ($j=A,B$)
\begin{equation}
\hat{\cal L}_j(\sigma)=\Gamma(2\hat j\sigma\hat j^\dag-\hat j^\dag\hat j\sigma-\sigma\hat j^\dag\hat j),
\end{equation}
which is the Liouvillian describing the damping of each cavity field due to a low-temperature bath (as is typical at optical frequencies) and acts on a generic density matrix $\sigma$. We now assume $\Gamma\gg|\Omega(x,y)|$, i.e., the bad-cavity limit, and trace out the cavity fields, so as to find an effective dissipative dynamical map for the qubits only, driven by a structured quantum environment that exhibits quantum correlations. Such an effective description is gathered following standard strategies for the derivation of adiabatically eliminated master equations~\cite{walls} as
\begin{equation}
\label{red}
\partial_{t}\rho_{12}=\textrm{Tr}_{AB}\left\{\hat{\cal L}_0\int^\infty_0e^{\hat{\cal L}t}\hat{\cal L}_0(\rho^{ss}_{AB}\otimes\rho_{12})dt\right\},
\end{equation}
which has been obtained  by defining a projection operator $\hat{\cal P}$ such that $\hat{\cal P}\rho_{12AB}=\rho^{ss}_{AB}\otimes\rho_{12}$ with $\rho^{ss}_{AB}$ the steady state of the cavity fields and $\rho_{12}$ the reduced density matrix of the atoms only. The projection operator is such that  $\hat{\cal P}\hat{\cal L}_0\hat{\cal P}\rho_{12AB}=0$, which is a key property for the derivation of Eq.~(\ref{red}). In Ref.~\cite{Paternostro}, it has been shown that the explicit form of the reduced master equation depends on  a Kossakowski matrix that is fully determined just by the second moments of the environmental modes. We introduce the {\it covariance matrix} ${\bm M}(\rho_{AB})$ with elements ${\bm M}_{ij}(\rho_{AB})=\textrm{Tr}[\rho_{AB}\{\hat{\bm q}_i,\hat{\bm q}_j\}/2]$, where $\hat{\bm q}=(\hat x_A\,\hat{p}_A\,\hat{x}_B\,\hat{p}_B)$ is the row vector of the two-mode field quadratures. Any two-mode covariance matrix can be transformed, by means of local symplectic operations, into the form~\cite{giedke}
\begin{equation}
{\bm M}=\begin{pmatrix}
{\bm n}&{\bm c}\\
{\bm c}^T&{\bm m}
\end{pmatrix},
\end{equation}
where ${\bm p}=p\openone_2~(p=n,m)$ account for the variances of each local mode and ${\bm c}=\textrm{diag}[c_1,c_2]$ describes the inter-mode correlations. In our formalism, the two-mode vacuum state corresponds to
$n=m=1/2$ with $c_{1,2}=0$. For a generic covariance matrix ${\bm M}$, the Kossakowski matrix that fully determines the dynamics of the two-atom system can be constructed as
\begin{equation}
{\bm K}=\gamma
\begin{pmatrix}
{\bm n}+i{\bm\Sigma}&{\bm c}\\
{\bm c}&{\bm m}+i{\bm\Sigma}
\end{pmatrix}
={\bm M}+i{\bm \Sigma}^{\oplus2}
\end{equation}
with $\gamma\propto|\Omega(x,y)|^2/\Gamma$ the effective two-qubit coupling rate and ${\bm \Sigma}=\begin{pmatrix}0&1\\
-1&0\end{pmatrix}$ the single-mode symplectic matrix~\cite{Paternostro,ACIP}. The reduced dynamics of the atoms is thus built as
\begin{equation}
\label{dynamo}
\partial_t\rho_{12}=\sum^4_{\alpha,\beta=1}{\bm K}_{\alpha\beta}(\hat{O}_{\alpha}\rho_{12}\hat{O}_\beta-\{\hat{O}_\beta\hat{O}_\alpha,\rho_{12}\}/2),
\end{equation}
where $\hat{O}_{\alpha}=\hat\sigma^1_\alpha\otimes\openone$ for $\alpha=1,2$ and $\hat{O}_{\alpha}=\openone\otimes\hat\sigma^2_{\alpha-2}$ for $\alpha=3,4$. Here, $\hat{\sigma}^j_{1(2)}$ is the
$x(y)$-Pauli matrix of qubit $j=1,2$. Complete positivity of the map originating from Eq.~(\ref{dynamo}) is ensured for ${\bm K}\ge0$, which is equivalent to the Heisenberg-Robertson uncertainty principle for the covariance matrix ${\bm M}$~\cite{walls}. In what follows, we will consider both the temporally-resolved two-atom dynamics achieved by solving Eq.~(\ref{dynamo}) and the steady-state one obtained by setting $\partial_t\rho_{12}=0$.
\begin{figure*}[t!]
\centering{(a)\hskip5cm(b)\hskip5cm(c)}
\includegraphics[width=0.28\linewidth]{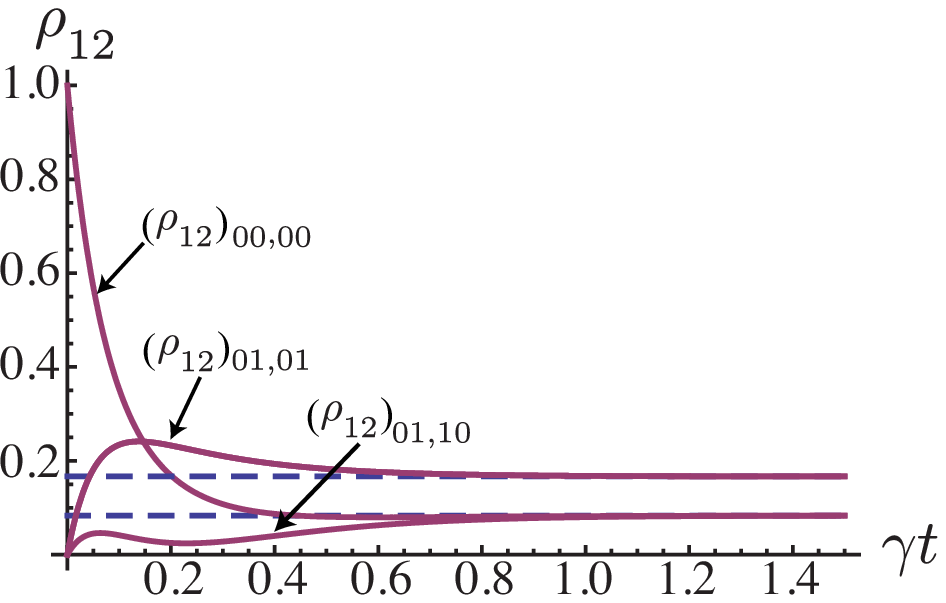}~~\includegraphics[width=0.28\linewidth]{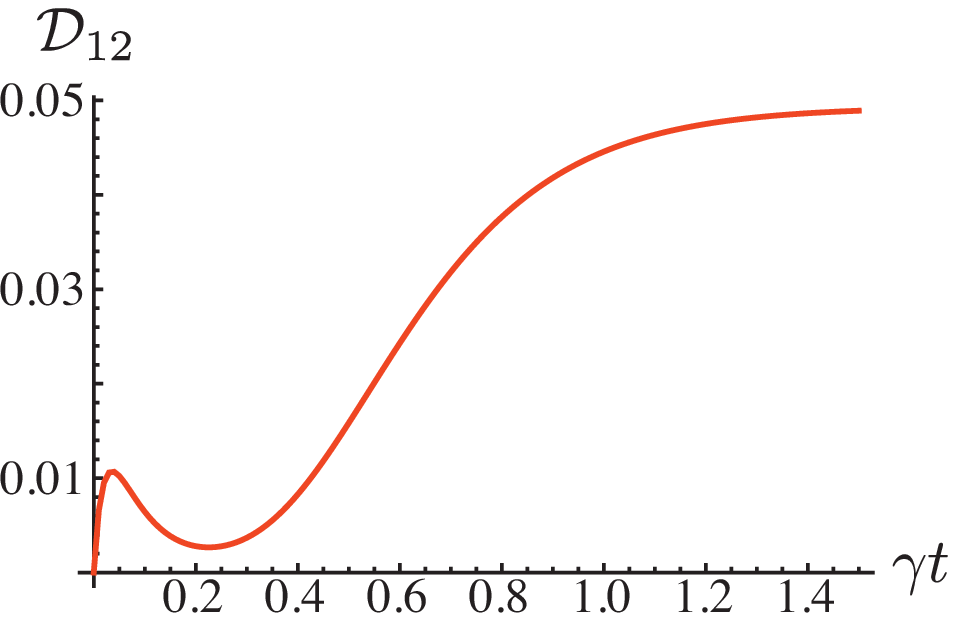}~~
\includegraphics[width=0.28\linewidth]{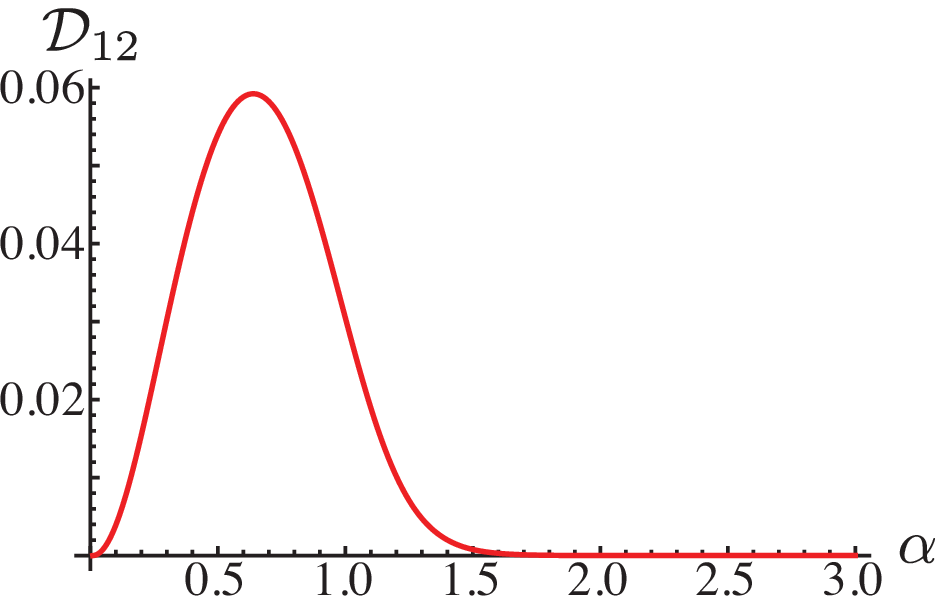}
\caption{(Color online) (a) Temporal behavior of the elements of the two-qubit density matrix for the case of an entangled single-photon driving field (solid lines). The dashed lines are the corresponding steady-state values. We have used the notation $(\rho_{12})_{ij,kl}=\langle{ij}|\rho_{12ss}|kl\rangle$. The initial state of the atoms is taken to be $\ket{00}_{12}$. (b) Quantum discord versus dimensionless interaction time $\gamma{t}$ for the case addressed in (a). (c) Atomic steady-state discord against the amplitude $\alpha\in\mathbb{R}$ in an ECS driving the entanglement-transfer process. Dynamically, the two-atom state is always separable with discord following a trend (at a set value of $\alpha$) very similar to what is shown in (b) for an entangled single-photon driving field. }
\label{fig5}
\end{figure*}

We start with the NOON driving field. Upon explicit calculation, it is straightforward to check that ${\bm M}(\rho_{\textrm{NOON}})=(N+1)\openone/2$ for any $N>1$ (here, $\rho_{\textrm{NOON}}=\ket{\textrm{NOON}}\bra{\textrm{NOON}}$). This simply implies that all the correlations in a NOON state with more than one photon are encoded in higher-order moments of the mode quadratures. As the process of entanglement transfer ruled by the interaction Hamiltonian $\hat H_{1A}+\hat H_{2B}$ relies on the second moments only, a point that has been duly stressed in Ref.~\cite{Paternostro2010}, it is clear that such a driving field does not embody a useful resource for the process, regardless of the dependence of the coupling strength $\Omega(x,y)$ on the details of the atomic transition across the cavities. In turn, this provides a rigorous explanation of the observations made earlier for the case of unitary dynamics. For $N=1$, on the other hand, the covariance matrix is nondiagonal and reads
\begin{equation}
{\bm M}(\rho_{\textrm{1001}})=\frac12
\begin{pmatrix}
2&0&1&0\\
0&2&0&1\\
1&0&2&0\\
0&1&0&2
\end{pmatrix}.
\end{equation}
This gives rise to the steady-state two-atom density matrix
\begin{equation}
\rho^{12}_{ss}=\frac{1}{12}
\begin{pmatrix}
1&0&0&0\\
0&2&1&0\\
0&1&2&0\\
0&0&0&7
\end{pmatrix},
\end{equation}
which has positive partial transpose, as it is straightforward to check, and thus describes a separable state. In Fig.~\ref{fig5}(a) we plot the time behavior towards steady state of the elements of the two-atom density matrix. Interestingly enough, not only is the two-atom state fully separable at all instants of time of the dynamics, it is also very weakly quantum correlated, in general. In order to support our claim, we have computed the entropic version of quantum discord~\cite{modi2012} proposed in Ref.~\cite{OZ} to test for more general quantum correlations seeded into the state of the two atoms. Quantum discord strives at capturing quantum correlations of a broad nature in a multipartite quantum system, witnessing the non-classicality of the way correlations are shared by the elements of (in general) a many-party register. In our case, a non-zero value of quantum discord in the bipartite atomic system at hand signals the existence of quantum features in the density matrix describing the state of atoms $1$ and $2$. The results are shown in Fig.~\ref{fig5}(b), where it is shown that a very small degree
of discord is shared by the atoms. The steady state, though, remains non-classically correlated, although separable.

We now consider an ECS driving field [cf. Eq.~(\ref{eq13})], whose covariance matrix is
\begin{widetext}
\begin{equation}
\label{cmECS}
{\bf M}(\rho_{\textrm{ECS}})={\cal N}_\alpha
\left(
\begin{array}{cccc}
{\alpha ^2+(1+e^{-\alpha ^2})(\alpha ^2+1)}& 0 &{e^{-\alpha ^2} \alpha ^2} & 0 \\
 0 &{e^{-\alpha ^2}(1-\alpha^2)+1}& 0 &{e^{-\alpha ^2} \alpha ^2} \\
{e^{-\alpha ^2} \alpha ^2} & 0 &{\alpha ^2+(1+e^{-\alpha ^2})(\alpha ^2+1)} & 0 \\
 0 &{e^{-\alpha ^2} \alpha ^2}& 0 &{e^{-\alpha ^2}(1-\alpha^2)+1}
\end{array}
\right).
\end{equation}
\end{widetext}
While both steady-state and time-resolved density matrices can be computed analytically using the formal apparatus described above, their expressions in terms of the
entries of the covariance matrix in Eq.~(\ref{cmECS}) are too lengthy to be reported here. We can then check for inseparability of the two-atom state, finding that, as in the entangled single-photon case, no entanglement is set by the dynamics, regardless of the form of the Rabi frequency $\Omega(x,y)$, both dynamically and at the steady state. As for the quantum discord, we have computed it for the steady-state density matrix against the amplitude $\alpha$ entering the driving field [see Fig.~\ref{fig5}(c)]: the discord is nonmonotonic against the ECS amplitude, showing that (very modest) non-classical correlations are set preferentially at small values of $\alpha$ (the maximum of the discord being found for $\alpha\simeq0.64)$. As $\alpha$ increases, the discord vanishes, thus giving back a state with only classical correlations, at most. In fact, for $\alpha\gg1$ the two-atom state is completely uncorrelated. Indeed, the off-diagonal terms in the steady-state density matrix go to zero as $\alpha$ increases, thus leaving a diagonal state with no correlations at all (actually, the atomic state tends towards a maximally mixed one).

The reason for the trends highlighted in this section is very clearly related to the fact that the covariance matrix of the field states driving the cavities fails to violate the criterion of positivity of partial transposition. Needless to say, as the field resources are non-Gaussian, this implies only that correlations are encoded in higher-order moments of the field's quadratures, as mentioned above. The dissipation-led entanglement-transfer process, though, strongly relies {\it only} on the mentioned second moments: as far as the process at hand is concerned, using the driving fields analyzed above is not different from driving the cavities with separable Gaussian states.
This analysis suggests the pathway that should be pursued, experimentally, when entangled single-photon and ECSs are used as resources: an almost unitary process needs to be in place, in order for the transfer process to actually take place. In contrast, should the transit of the atom occur in times comparable to the cavity lifetime (alternatively, should the quality of the cavities be not high enough), the scheme will fail, leaving, at its best, atomic states that are only weakly quantum correlated.

\section{Conclusions}
\label{Con}
We have investigated the entanglement transfer between flying qubits and optical fields based on a cavity-QED system. We have considered two typical quantum-correlated driving fields, NOON and entangled coherent states, and found that entangled atomic qubits can be prepared, effectively, regardless of the details of their motion. We have discussed diverse aspects of our system, including its robustness to the effects of non-negligible cavity dissipation and sensitivity to the value of the Rabi frequency, finding that the maximum entanglement the qubits can achieve is not determined by their motion but by the characteristics of the cavity fields, which in turn depend on the external driving and cavity damping rate. 
The high degree of robustness of the scheme (for experimentally realistic parameters) allows consideration of the consecutive passage of pairs of independent qubits through the entangled cavities and the construction of a stream of entangled qubits. The potential and limitations of such scheme will be addressed elsewhere.

\acknowledgments
J.L. is grateful to the Institute of Opto-Electronics, Shanxi University, for kind hospitality during the completion of this work, and to the School of Mathematics and Physics, Queen's University Belfast, for financial support from a Purser Studentship. This work was supported by the State Basic Key Research Program of China (Grant No. 2012CB921601), the China National Funds for Distinguished Young Scientists (Grant No. 11125418), and the National Natural Science Foundation of China (Grants No. 10974125 and No. 61121064). M.P. thanks the U.K. EPSRC for financial support through a Career Acceleration Fellowship and the ``New Directions for EPSRC Research Leaders" initiative (Grant No. EP/G004759/1).

\end{document}